\documentclass[a4paper,10pt,reqno,superscriptaddress]{revtex4}

\draft 

\usepackage[centertags]{amsmath}
\usepackage{amsfonts}
\usepackage{amssymb}
\usepackage{amsthm}
\usepackage{newlfont}
\usepackage{stmaryrd}
\usepackage{mathrsfs}
\usepackage{euscript}

\usepackage{color}

\usepackage{dcolumn}

\theoremstyle{plain}

\theoremstyle{definition}

\theoremstyle{remark}

\let\al=\alpha   \let\ep=\epsilon

\let\De=\Delta   \let\Om=\Omega

\newcommand{\caO}{{\mathcal O}}

\newcommand{\opunit}{\text{1}\kern-0.22em\text{l}}

\DeclareMathAlphabet{\mathpzc}{OT1}{pzc}{m}{it}

\newcommand{\id}{\textrm{d}}

\begin{document}

\title{General no-go condition for stochastic pumping}

\author{Christian Maes}
\homepage[]{itf.fys.kuleuven.be/~christ/}
\affiliation{Instituut voor Theoretische Fysica, K.U.Leuven, Belgium}
\author{Karel Neto\v{c}n\'{y}}
\email{netocny@fzu.cz}
\affiliation{Institute of Physics AS CR, Prague, Czech Republic}
\author{Simi R.\ Thomas}
\affiliation{Instituut voor Theoretische Fysica, K.U.Leuven, Belgium} 

\begin{abstract}
The control of chemical dynamics requires understanding the effect
of time-dependent transition rates between states of
chemo-mechanical molecular configurations.  Pumping refers to
generating a net current, e.g. per period in the time-dependence,
through a cycle of consecutive states. The working of artificial
machines or synthesized molecular motors depends on it. In this
paper we give short and simple  proofs of no-go theorems, some of
which appeared before but here with essential extensions to
non-Markovian dynamics, including the study of the diffusion
limit. It allows to exclude certain protocols in the working of chemical motors where only the depth of the energy well is changed in time and not the barrier height between pairs of states.  We also show how pre-existing steady state currents are in general modified with a multiplicative factor when this time-dependence is turned on.
\end{abstract}

\maketitle

\section{Introduction}

Molecular cybernetics deals with the control and the steering of
system of molecules.  As part of systems chemistry it investigates models of  chemical networks with time-dependent
dynamics.  A common application is found in the study of molecular
motors or mesoscopic machines. Here, chemical or
electrical gradients repetitively and progressively drive a system
away from equilibrium in such a way that when the motor returns to its
original configuration, a physical task performed by the machine
is not undone~\cite{tsong02,ferin06}.
This technology is ubiquitous in nature in the form of translational and rotational movements, e.g.\ in muscle fibres, bacterial flagella and
cilia~\cite{schil,vale00} or to transport material within and across a cell membrane, either powered by solar energy (as in photosynthesis) or by chemical
energy stored in molecular bonds (e.g.\ ATP).

Artificial designs  analogous to such motors use
external time-dependent perturbations
such as light, heat or chemical stimulus to drive the
system~\cite{leigh03,leigh04,klok08,kelly99,feringa99}. It is thus
relevant, in attempts to synthesize and control
artificial molecular motors, to understand the relation between
the external pumping and creation of
systematic flows.
Previously, similar questions were asked by Thouless~\cite{thoul83}, for electronic pumping. Astumian and
Der\'enyi~\cite{astu01} studied charge transfer
from a lower to a higher chemical potential by varying the gate
and portal energies.
Astumian also analyzed the adiabatic regime of ion pumping in
externally driven protein structures~\cite{astu03} and in a
molecular motor based on a three-ring catenane~\cite{astu07}. A
general theory of adiabatic pumps in terms of geometrical phase
was proposed by Sinitsyn and Nemenman~\cite{sine07}.
Chernyak and Sinitsyn~\cite{cher09} have discovered that the adiabatic pumping currents become quantized at low temperatures.
Generalizations beyond adiabatic regimes are so far limited to Markov models.\\

In the present paper we concentrate on no-go or
no-pumping theorems, stating the absence of a net time-averaged
current under certain protocols.
We refer to the experiment by Leigh et al.~\cite{leigh03} on unidirectional motion in (2)- and (3)-catenanes.
Rahav, Horowitz and Jarzynski~\cite{jar08} were the first to give a
no-pumping theorem for jump processes with non-adiabatic pumping and generalization to diffusion processes~\cite{jar09}. This was further studied and systematized by Chernyak and Sinitsyn~\cite{cher08} so that
the following general conclusion was reached: when the dynamics
can be modeled as a Markov state system for which the transitions
between states $x$ and $y$ have an Arrhenius-type time-dependence
\[
  w_t(x,y) = A(x,y)\,e^{\frac{G_t(x)}{k_B T}}, \quad A(x,y) = A(y,x)
\]
with periodic time-dependent energy wells $G_t(x)$ and with constant energy barriers as represented by the symmetric factors $A(x,y)$,
then the time-averaged current $J(x,y)$ along every transition
$x\rightarrow y$
is zero. As a result, no net work can be done with such protocol.

The purpose of the paper is to extend the arguments proposed in
Ref.~\onlinecite{cher08} and to give the shortest general proof of
this result (Section~\ref{pro}), which at the same time also
applies to classes of non-Markov models (Section~\ref{semi}) and
for which the diffusion limit becomes equally simple
(Section~\ref{dif}). At the end we put the result into a broader
context by showing that the time-dependent protocol under
consideration in arbitrary (in general nonequilibrium) systems
modifies all currents by a global multiplicative factor
(Section~\ref{mod}).
We start with the general set-up in terms of a Markov jump process.

\section{Set-up}\label{setup}

Markov state models and their extensions are important tools for modeling thermodynamic processes of open systems~\cite{keizer} and they find numerous applications in chemical kinetics~\cite{keizer,park06,sid05,greg09} and in bio-chemistry~\cite{kurzy}.\\

In our abstract framework we use $x,y,\ldots$ to denote the
long-lived or metastable states that locally minimize a given free
energy landscape $G_\alpha(x)$ under equilibrium conditions; for instance,
they refer to chemo-mechanical configuration of molecules
within a suitable coarse-grained level of description.
Together, they form the vertices of a stochastic network with
bonds between pairs $(xy)$ indicating possible transitions. The
dynamics is encoded in the transition rates $w_\alpha(x,y)$
satisfying the condition of detailed balance
\begin{equation}\label{adb}
  e^{-G_\alpha(x)}\,w_\alpha(x,y) = e^{-G_\alpha(y)}\,w_\alpha(y,x)
\end{equation}
with the free energies $G_\al$ in $k_B T$ units.
The index $\al$ indicates a possible dependence on an externally controlled parameter which varies the depth of the local free
energy minima $G_\alpha(x)$. Detailed balance~\eqref{adb} implies that for
fixed $\alpha$ the dynamics (by assumption ergodic) relaxes to the equilibrium Gibbs distribution $\propto \exp [-G_\alpha(x)]$.

As indicated before, we restrict ourselves to a protocol that makes the transition rates,
\begin{equation}\label{cr}
  w_t(x,y) =  A(x,y)\,e^{ G_{\al(t)}(x)}\,,\quad
  A(x,y) = e^{-\De(x,y)}
\end{equation}
time-dependent via the parameter $\alpha = \alpha(t)$
in the free energy minima $G_\al$.
$\De(x,y) = \De(y,x)$ are the effective barrier heights which are kept fixed.
In general, $A(x,y)$ just specifies the
Arrhenius prefactor in the transition rates
and $\De(x,y)$ deviates
from a true barrier energy by geometrical corrections. However,
a good approximation in many applications is
to consider $G(x)$ and $\De(x,y)$ as independent in the sense that they can be manipulated
independently from each other by suitably varying the minima and the barrier heights respectively.\\

It has been shown in~\cite{jar08} that no energy pumping is possible for such systems in the
following sense. Let $\rho_t$ be the instantaneous distribution
function and let $j_t(x,y)$ be the corresponding
instantaneous mean current between pairs of states,
\begin{equation}\label{eq: jt}
  j_t(x,y) = \rho_t(x)\,w_t(x,y) - \rho_t(y)\,w_t(y,x)
\end{equation}
as obtained from the Master equation written in the form
\begin{equation}\label{eq: master}
  \frac{d\rho_t(x)}{dt} + \sum_{y} j_t(x,y) = 0
\end{equation}
There is no strictly stationary distribution as the rates are
time-dependent, but when the protocol $\alpha(t)$ is periodic in
time, we expect to find that $\rho_t$  itself becomes periodic in
time, at least for sufficiently large times $t$. In any event, we
can define the time-averaged current
\begin{equation}\label{cur}
  J(x,y)=  \lim_{T}\frac{1}{T} \int_0^T j_t(x,y)\,\id t
\end{equation}
and the no-pumping theorem states that this long-time average
equals zero for all pairs of states $x,y$.
Thinking of independent particles hopping
on the network over $x\rightarrow y$ with rate $w_t(x,y)$, the
no-pumping refers to having no net
time-averaged flow of particles between any two nodes $x$ and $y$.\\

In the next section we present a simple derivation of this result
within the present set-up of Markov state models. Later sections
will provide extensions to the semi-Markov and the diffusion
systems, giving the main results of this paper.

\section{No-pumping theorem}\label{pro}

Before starting with the proof we remind the reader of two important facts. First of all, the no-pumping theorem is only valid for {\it some specific types}
of time-dependence ---
in general for those considered in~\eqref{cr}.  Even when the rates are satisfying the condition of detailed
balance \eqref{adb} for each fixed value of the parameter
$\alpha$, there is no {\it a priori} reason why there could not arise
a net current $J(x,y)$ in the process with time-dependent $\alpha(t)$.
In fact, that is exactly what happens in so called flashing (and
other) ratchets where the change in the potential landscape
produces a net flow of particles~\cite{reim01}.
For example, a system (like a ratchet) with transition rates
\[
  \tilde{w}_t(x,y)= A(x,y)\,e^{-\frac{1}{2}[G_t(y)-G_t(x)]},\quad
  A(x,y) = A(y,x)
\]
also satisfies detailed balance \eqref{adb} for each fixed time $t$ and can be
written analogous to \eqref{cr} as
\begin{equation*}
  \tilde{w}_t(x,y) = e^{G_t(x) - \tilde{\De}_t(x,y)}
\end{equation*}
but the effective barriers
$\tilde \De_t(x,y) = [G_t(x) + G_t(y)]/2 - \ln A(x,y)$ have become  \emph{time-dependent}. Within the framework of the no-go theorem there is absolutely no reason now that the net currents would be
identically zero (unless further symmetries are imposed).\\

Secondly, the geometry of the stochastic network is certainly
relevant for the possible generation of a current. In fact, the
net current $\int_0^T j_t(x,y)\,\id t$ over any edge $(xy)$
connecting two otherwise disconnected subgraphs is a total
time-difference of the form $N_T(x,y)-N_0(x,y)$ and hence
automatically approaches zero when time-averaged as in
\eqref{cur}. Thus $w_t(x,y)$ can be arbitrary (= no restriction)
over such a ``bridge'' and the restricted form of time-dependence
as in~\eqref{cr} is only required over
those edges $(xy)$ which belong to a loop.\\

We now come to our formulation of the no-pumping theorem. Consider
the class of Markov jump processes with states
$x,y,\ldots$ as in Section~\ref{setup}. For all bonds $(xy)$
in our stochastic network
that are part of a loop in the network we require that the
time-dependence in the transition rates is of the form
\begin{equation}\label{para}
  w_t(x,y)= \lambda_t(x)\,p(x,y)
\end{equation}
where $\lambda_t(x) = \sum_y w_t(x,y)$ is the time-dependent
escape rate and $p(x,y)$ is a time-independent transition
probability; $p(x,y) \geq 0, \sum_y p(x,y) =1$. We assume that the
matrix $[p(x,y)]$ is irreducible so that there is a unique left
eigenvector $\mu$ for eigenvalue 1: $\sum_x \mu(x) p(x,y) =
\mu(y)$. (That is automatically so when the network of states is
connected via $p(x,y) > 0$ --- Perron-Frobenius theorem.)   We also
assume detailed balance, i.e., for some potential $V$,
\begin{equation}\label{debe}
 e^{-V(x)}\,p(x,y) = e^{-V(y)}\,p(y,x)
\end{equation}
so that in fact $\mu(x) \propto e^{-V(x)}$.
Finally, we suppose that the limit
\begin{equation}
  \varpi(x) =  \lim_{T}\frac{1}{T} \int_0^T
  \rho_t(x)\,\lambda_t(x)\,\id t
\end{equation}
exists. (That is automatically satisfied when the time-dependence
is periodic but, clearly that is not strictly necessary.)

The no-pumping theorem is now easily proven as follows. Since the
full time evolution is obtained by solving~\eqref{eq:
jt}--\eqref{eq: master}, the time integral of the latter gives
\begin{equation}\label{6b}
  0 = \lim_{T}\frac{1}{T} \int_0^T \frac{\id \rho_t}{\id t}\,\id t
  = -\sum_{y} J(x,y)
\end{equation}
and
\begin{equation}\label{6}
  J(x,y)= \varpi(x)\,p(x,y)- \varpi(y)\,p(y,x)
\end{equation}
where we have inserted condition~\eqref{para}. Hence,
$\sum_x\varpi(x)\,p(x,y) = \varpi(y)$, which is the
stationary Master equation for a time-independent Markov chain
with (unnormalized) distribution $\varpi$ and current $J$. By irreducibility and by detailed balance~\eqref{debe}, the equations~\eqref{6b}--\eqref{6} have the unique solution
\begin{equation}\label{6c}
  \varpi(x) \propto e^{-V(x)},\qquad
  J(x,y) = 0 \text{ (all pairs)}
\end{equation}
as was to be proven.\\

The conditions~\eqref{para}--\eqref{debe} are just equivalent to~\eqref{adb}--\eqref{cr} and, hence,
the above theorem can immediately be applied to the situation of
Section~\ref{setup}. Specifically, for a system with
free energy wells $G_\al(x)$ and effective energy barriers $\De(x,y) = \De(y,x)$ as in~\eqref{cr},
we consider an arbitrary cyclic path $\al(t) = \al(t + T)$ with
some period $T$. Keeping the energy barriers constant, the
condition~\eqref{para} is verified with time-dependent escape
rates
\[
  \lambda_t(x)= e^{G_{\al(t)}(x)}\,\sum_y e^{-\De(x,y)}
\]
and time-independent transition probabilities
\[
  p(x,y) = \frac{e^{-\De(x,y)}}{\sum_y e^{-\De(x,y)}}
\]
satisfying detailed balance~\eqref{debe} for
$V(x)= -\ln \sum_y e^{-\De(x,y)}$.
\\

As a summary, the essential ingredients are two-fold.  First, the original time-dependent jump process
is detailed balanced for each fixed $t$, and secondly, the transition rates can be decomposed into a product of time-dependent escape rates $\lambda_t(x)$ and time-independent transition probabilities $p(x,y)$.

The idea of the proof above, at least for a periodic
time-dependence, is that the time-averaged current in the original
system \eqref{cr} exactly coincides with the stationary current in a
temporally coarse-grained system with transition probabilities
$p(x,y)$.  As the original process is detailed balanced for each
fixed time,  the stationary coarse-grained process is
time-reversal symmetric and therefore the net current in the
original process vanishes.\\

As a final comment, it is important to realize that
the vanishing of the
net (i.e., time-averaged) current, $J(x,y) = 0$, does \emph{not} imply that $j_t(x,y) = 0$ at each time, unless the process runs in the
quasistatic regime.
Related to that, the overall dissipation does remain
nonzero in general. For this it suffices to look at the time-averaged entropy flux
\[
  \text{EF} \equiv \frac 1{2T}  \sum_{x,y}\int_0^T j_t(x,y)\, \ln
  \frac{w_{\alpha(t)}(x,y)}{w_{\alpha(t)}(y,x)}\,\id t \geq  0
\]
We refer e.g.\ to Section~3 in Ref.~\onlinecite{mnw} for putting this
expression for the entropy flux in a thermodynamic context.
Inserting \eqref{adb} we compute
\begin{equation}\label{ef}
  \text{EF} = \frac{1}{T} \int_0^T
  \Bigl\langle \frac{\id G_{\alpha(t)}}{\id t}
  \Bigr\rangle_{\rho_t} \id t
\end{equation}
over the average work
$\langle \id G_{\alpha(t)}/\id t \rangle_{\rho_{t}}
\equiv \sum_x \rho_{t}(x)\,\id G_{\alpha(t)}(x)/\id t$.
Only in the quasi-static (or adiabatic)
limit when we can take
$\rho_t(x) \propto \exp [-G_{\alpha(t)}(x)]$
do we get that
$\langle \id G_{\alpha(t)}/\id t\rangle_{\rho_{t}}
= - \id/\id t\, \ln \sum_x \exp [-G_{\alpha(t)}(x)]$
and is the entropy flux $\text{EF} = 0$.

\section{Non-Markov generalization}\label{semi}

By the simplicity of the proof above the arguments allow a
natural extension to more general jump processes including those
with non-exponential waiting time distributions. These are called
semi-Markov systems or continuous-time random walks, here on the
chemical network. The fact that the presented method survives here is important, especially since many biophysical and biochemical processes are believed to be essentially non-Markovian for a natural choice of states~\cite{fulin98}.\\

We consider a jump process for which the main change with respect
to the Markov case consists in its dependence on the time $t_0$
of the previous jump.
In that way, given that the system is in
state $x$ at time $t$ since its last jump to $x$ at $t_0$, the
probability that the next jump occurs within the time-interval
$[t,t+ \id t]$ is given by $\lambda(x;t_0,t)\,\id t$.  (The Markov case
corresponds to $\lambda(x;t_0,t)=\lambda_t(x)$.) The probability
rate that the next jump goes to $y$ is then
\begin{equation}
  w(x,t_0;y,t)= \lambda(x;t_0,t)\,p(x,y)
\end{equation}
generalizing the time-dependent Markov transition rates
\eqref{para}.  We keep the same assumptions on the
transition matrix $[p(x,y)]$, with its most important property being the condition of detailed balance \eqref{debe}. The $p(x,y)$ define what is often called the embedded Markov chain. The complication of the memory present in the escape rates $\lambda(x;t_0,t)$ turns out to be irrelevant for our proof of the no-pumping, as we now show.\\

The probability density that at time $t$ the system is found in state $x$ and that the last jump before $t$ occurred within $[t_0,t_0+ \id t_0]$ is denoted by $\rho(x;t_0,t)\,\id t_0$ --- it relates to the standard single-time distribution as
\[
\rho_t(x)=\int_{0}^{t} \rho(x;t_0,t)\, \id t_0
\]
The mean current $j(x,t_0;y,t)\,\id t_0$ counts the expected rate of (directed) jumps $x\rightleftarrows y$ at time $t$ when the previous jump occurred in $[t_0,t_0+\id t_0]$:
\[
j(x,t_0;y,t)= \rho(x;t_0,t)\,w(x,t_0;y,t) - \rho(y;t_0,t)\,w(y,t_0;x,t)
\]
It is related to the standard mean current as
\begin{equation}\label{sta}
j_t(x,y)= \int_{0}^{t} j(x,t_0;y,t) \, \id t_0
\end{equation}
By construction, the single-time quantities $\rho_t$ and $j_t$ satisfy the balance equation
\begin{equation}\label{mas}
 \frac{\id\rho_t(x)}{\id t} +\sum_y j_t(x,y)= 0
\end{equation}
We proceed analogously as in the Markov case.  We assume that the limiting quantities
\begin{align}
 \varpi(x) &= \lim_{T} \frac{1}{T} \int_{0}^{T} \int_{0}^{t} \rho(x;t_0,t)\lambda(x;t_0,t) \,\id t_0 \,\id t
\\
  J(x,y) &= \lim_{T} \frac{1}{T} \int_{0}^{T} j_t(x,y)\, \id t
\end{align}
are well defined. Then again,
\begin{equation}
J(x,y) = \varpi(x)p(x,y)-\varpi(y)p(y,x)
\end{equation}
and from integrating \eqref{mas},
 \begin{equation}
\sum_{y} J(x,y)=0
\end{equation}
By detailed balance \eqref{debe} we reach the conclusion $J(x,y)=0$ which ends the proof.  As before,
\textit{time-homogeneity and detailed balance of the embedded Markov chain imply that the net flux through any pair $(xy)$ asymptotically goes to zero.}

\section{Diffusion limit}\label{dif}

The same ideas apply to diffusion processes.  We can obtain them
as continuum limits of Markov jump processes. For simplicity we
assume here that the stochastic network is a one-dimensional
chain, possibly turned into a circle.  The lattice spacing is
denoted by $\epsilon$ and we will take the continuum limit
$\epsilon \to 0$. We think of the transitions as the hopping of a
random walker between nearest neighbor sites
$x\rightarrow x\pm \epsilon$ with rates depending on space- and time-dependent
amplitudes $D_t(x)$ and a potential $U_t(x)$:
\begin{equation}\label{neigh}
\begin{split}
 w^\epsilon&(x,x\pm\epsilon) = \epsilon^{-2}\,\sqrt{D_t(x)D_t(x\pm\epsilon)}\,
 e^{[U_t(x)-U_t(x\pm\epsilon)]/2}
\\
 &= \epsilon^{-2}D_t(x)\pm \frac{1}{2}\,\epsilon^{-1}D_t(x)\,[\ln{D_t}-U_t]'(x)
 +{\cal O}(1)
\end{split}
\end{equation}
by expanding
around $\epsilon=0$ and with $\caO(1)$ indicating an error term
that remains bounded along $\ep \to 0$.  As in
\eqref{para} we decompose these transition rates
$w^\epsilon(x,x\pm \epsilon)$ into a product of escape rates
$\lambda^{\epsilon}_t(x)$ and transition probabilities
$p^\epsilon(x,x\pm \epsilon)$, to find
\[
 \lambda^\epsilon_t(x)=2\epsilon^{-2}D_t(x)+{\cal O}(1)
\]
and
\begin{equation}\label{peps}
\begin{split}
 \nonumber p_\epsilon(x,x\pm\epsilon) &=
 \frac{1}{2}\pm \frac{\ep}{4}\,[\ln{D_t}-U_t]'(x)+{\cal O}(1)
\\
 &= \frac{1}{2}\, e^{[V(x)-V(x\pm\epsilon)]/2} + {\cal O}(1)
\end{split}
\end{equation}
under the condition that $V = U_t - \ln D_t $ does not depend on
time. That  then reproduces the form~\eqref{para} under which the
no-pumping holds.  This comparison to jump processes
predicts under what natural condition we may expect zero net
current also for diffusions.
Next we give a direct argument that confirms this idea.\\

We are now dealing with a Langevin-type equation in It\^o
form
\begin{equation}\label{ord}
  \id x_t = -D_t(x_t)\,U'_t(x_t)\,\id t + D'_t(x_t)\,\id t
  + \sqrt{2D_t(x_t)}\,\id B_t
\end{equation}
where $\id B_t/\id t$ indicates standard white noise. That
describes an overdamped one-dimensional diffusion on a ring (for
periodic boundary condition) or on the line, in a time-dependent
potential landscape $U_t(x)$ and with time-dependent diffusion
parameter $D_t(x)$ (setting the environment temperature equal to
one).

The corresponding Fokker-Planck equation reads
\begin{equation}\label{fk}
\begin{split}
  &\hspace{10mm} \frac{\partial \rho_t(x)}{\partial t} + j'_t(x) = 0
\\
  &j_t(x) = -\rho_t(x)\,D_t(x)\,U'(x) - D_t(x)\,\rho'_t(x)
\end{split}
\end{equation}
The same ideas as above can now be applied.  As before we define the quantities
\begin{equation}
 \varpi(x)=\lim_{T} \frac{1}{T} \int_{0}^{T} \rho_t(x)D_t(x) \,\id t
\end{equation}
and the time-averaged current
\begin{equation}\label{ac}
  J(x)=\lim_{T} \frac{1}{T} \int_{0}^{T} j_t(x) \,\id t
\end{equation}
By time-integrating~\eqref{fk} we reach the stationarity condition
\begin{equation}
J'(x)=0
\end{equation}
For ``effective'' potential $V=U_t-\ln D_t$ and assuming that $V$ is time-independent, one gets
\begin{equation}
\begin{split}
  J(x) &= \lim_{T}\frac{1}{T} \int_{0}^{T} [-\rho_t\, D_t\,
  V'(x)-(\rho_t\, D_t)'] \,\id t
\\
  &= -\varpi(x)V'(x)-\varpi'(x)
\end{split}
\end{equation}
As a consequence, the time-averaged characteristics of the original process~\eqref{ord} coincide with the stationary characteristics of a time-independent detailed balanced diffusion in the potential landscape $V$ with unit diffusion parameter.  Therefore,
$\varpi(x) \propto \exp [-V(x)]$ and $J = 0$ is the (unique) solution.\\

We conclude that for diffusion process with only gradient forces the condition
\begin{equation}\label{cond}
  \frac{\partial}{\partial t} \,(D_t\,e^{-U_t}) = 0
\end{equation}
is sufficient for the vanishing of the average current
\eqref{ac}. We recognize in \eqref{cond} the condition for
time-independence of the transition probability $p^\epsilon(x,x\pm
\epsilon)$  of the random walk in \eqref{peps}.
Our result also agrees with the no-pumping condition found
previously by a more complicated method~\cite{jar09}.
A no-pumping theorem for a general Langevin process on a compact manifold of arbitrary dimension was given in Ref.~\onlinecite{cher08}, using a method similar to the above.

\section{Nonequilibrium generalization}\label{mod}

A natural question arises how the extra time-dependence affecting
only the energy wells or escape rates like in~\eqref{para}
modifies the long-time characteristics of a general nonequilibrium
system with steady currents already present. We answer here
that question by generalizing the above argument, restricting
ourselves to the case of jump processes (Section~\ref{pro}). This
will throw more light into the nature and robustness of no-pumping theorems.\\

As before in~\eqref{cr} we start from transition rates
\begin{equation}
  w_t(x,y) = w(x,y)\,e^{G_t(x)}
\end{equation}
with time-dependent energy function $G_t(x)= G_{\alpha(t)}(x)$ but
now we do \emph{not} assume detailed balance for the reference process with rate $w(x,y)$. That reference could thus very well
correspond to a driven nonequilibrium but time-homogeneous
system. Let $\rho^s(x)$ be the stationary density of the
reference, with steady currents $j^s(x,y)= w(x,y)\,\rho^s(x) -
\rho^s(y)\,w(y,x)$, $\sum_y j^s(x,y) = 0$.

Inspecting formulas~\eqref{6b}--\eqref{6c}, we find that the long-time averaged current is
\begin{equation}\label{cac}
\begin{split}
  J(x,y) &= \lim_T \frac{1}{T} \int_0^T j_t(x,y)\,\id t
\\
  &= \varpi(x)\,w(x,y) - \varpi(y)\,w(y,x)
\end{split}
\end{equation}
with
\[
  \varpi(x) = \lim_T \frac{1}{T} \int_0^T \rho_t(x)\,e^{G_t(x)}\,\id t
\]
and satisfying the stationarity condition in all nodes:
\begin{equation}\label{cac1}
  \sum_y J(x,y) = 0
\end{equation}
By the assumed ergodicity, the equations
\eqref{cac}--\eqref{cac1} have a unique solution in the form
\begin{equation}
  \varpi(x) = \Om\,\rho^s(x)\,,\quad
  J(x,y) = \Om\,j^s(x,y)
\end{equation}
with the normalization
\begin{equation}\label{eq: adiabatic}
  \Om = \lim_T \frac{1}{T} \sum_x \int_0^T \rho_t(x)\,e^{G_t(x)}\,\id t
\end{equation}
Hence, we have arrived at an important conclusion: For the time-dependent protocols under consideration, the time-average current~\eqref{cac} is merely a \emph{global multiplicative factor} of the reference steady current. If the latter is zero, there is also no resulting pumped current and we recover the original results.

Remark that the ergodicity assumption is not essential: In general, the dynamics decomposes into several ergodic components made of mutually connected states. Within each of the components, the reference stationary distribution $\rho^s$ is unique and the above argument readily applies, up to that the current multiplicative factor $\Om$ is now generally different for each component and it depends on the initial distribution $\rho_0$.\\

In the case of a quasi-static process the system passes through
the states
\begin{equation}
  \rho_t(x) = \frac{1}{Z_t}\,\rho^s(x)\,e^{-G_t(x)}\,,\quad
  Z_t = \sum_x \rho^s(x)\,e^{-G_t(x)}
\end{equation}
which are stationary with respect to the instantaneous energy
landscape $G_t$. Indeed, these distributions correspond  to the
currents $j_t(x,y) = j^s(x,y) / Z_t$ such that $\sum_y j_t(x,y) =
0$. The normalization factor~\eqref{eq: adiabatic} then becomes
\begin{equation}
  \Om = \lim_T \frac{1}{T} \int_0^T \frac{\id t}{Z_t}
\end{equation}
One also checks that the total dissipation as measured by the
time-integrated outgoing entropy flux \eqref{ef} gets modified by
the same factor $\Om$ (but only) in the adiabatic limit.

\section{Conclusion}
The no-pumping theorem relies on the decomposition of the
jump rates into a time-dependent escape rate and a
time-homogeneous stochastic matrix (embedded Markov chain). Then,
time-integrated currents depend on the given time-dependent
protocol only via a global multiplicative factor.  When the
embedded Markov chain satisfies detailed balance, the no-pumping
result appears.

While the temporal coarse-graining can indeed express
the original time-integrated current in terms of the  steady
current for the embedded Markov chain, that is not true for the
dissipation or entropy flux which remains nonzero even in case the
embedded Markov chain is detailed balanced. The same ideas and
methods of proof apply to semi-Markov processes and to the
diffusion limit.

\begin{acknowledgements}
The authors thank Chris Jarzynski for initial discussions. CM\
benefits from the Belgian Interuniversity Attraction Poles
Programme P6/02.  KN\ acknowledges the support from the project
AV0Z10100520 in the Academy of Sciences of the Czech Republic and
from the Grant Agency of the Czech Republic (Grant no.~202/07/J051). The work was started by CM\ and KN\ at the Yukawa Institute for Theoretical Physics in Kyoto, while attending the YKIS 2009 ``Frontiers in Nonequilibrium Physics.''
\end{acknowledgements}


\end{document}